\documentstyle[12pt]{article}
\def\boxriv#1#2#3#4#5{}
\newskip\humongous \humongous=0pt plus 1000pt minus 1000pt
  \newif\ifdtup

\def\ltap{\raisebox{-.4ex}{\rlap{$\sim$}} \raisebox{.4ex}{$<$}}

\def\frac#1#2{ {{#1} \over {#2} }}

\def\abar{{\bar \alpha_S}}
\def\abn{{{\bar \alpha_S} \over N}}
\def\as{\alpha_S}

\def\smsbar{\overline{\mbox{\scriptsize MS}}}
\def\msbar{\overline{\mbox {\rm MS}}}

\def\ga{\gamma}
\def\si{\sigma}

\def\rt){\right)}
\def\lt({\left(}
\def\rq]{\right]}
\def\lq[{\left[}

\def\figcap{\section*{Figure Captions\markboth
        {FIGURECAPTIONS}{FIGURECAPTIONS}}\list
        {Figure \arabic{enumi}:\hfill}{\settowidth\labelwidth{Figure
999:}
        \leftmargin\labelwidth
        \advance\leftmargin\labelsep\usecounter{enumi}}}
 \relax

\def\np#1#2#3{Nucl.\ Phys.\ B#1 (19#3) #2}
\def\pl#1#2#3{Phys.\ Lett.\ #1B (19#3) #2}

\def\prep#1#2#3{Phys.\ Rep.\ #1 (19#3) #2}

\catcode`\@=11

\def\ps@headings{\def\@oddfoot{}\def\@evenfoot{}
\def\@oddhead{\hbox{}\hfill --\thepage{}-- \hfill}
\def\@evenhead{\@oddhead}
\def\subsectionmark##1{\markboth{##1}{}}
}

\ps@headings

\catcode`\@=12

\relax

\relax

\begin{document}

\begin{titlepage}
\renewcommand{\thefootnote}{\fnsymbol{footnote}}
\begin{flushright}
     Cavendish-HEP-94/08 \\   July 1994
     \end{flushright}
\vspace*{5mm}
\begin{center}
{\Large \bf \boldmath Next-to-leading Corrections  \\
 at Small $x$  from Quark Evolution\footnote{Talk given at
the  XXIX Rencontres de Moriond, 19-26 March 1994, Meribel, France,
to appear in the Proceedings.}}\\
\vspace*{1cm}

        \par \vskip 5mm \noindent
        {\bf F. Hautmann}\\
        \par \vskip 3mm \noindent
        Cavendish Laboratory\\
        Department of Physics, University of Cambridge\\
        Madingley Road, Cambridge CB3 0HE, UK\\

\par \vskip 1cm

\end{center}
\vspace*{1cm}

\begin{center} {\large \bf Abstract} \end{center}
\begin{quote}
Deep inelastic processes at small $x$ are discussed in the framework
of perturbative QCD at high energy.
New results are presented on the
quark anomalous dimensions beyond the leading logarithmic approximation,
 and their relevance
to the structure functions being  measured at HERA is pointed out.
\end{quote}
\vspace*{2cm}
\end{titlepage}

\noindent {\bf 1. Introduction   }
\vskip 0.1 true cm

 Quantitative tests of QCD and searches for new physics at
present and future hadron colliders are  carried out
at an increasingly large
centre-of-mass energy $\sqrt{S}$. In this regime a new kinematic region
opens up, characterized by small values of
 the ratio $x = Q^2/S \,$
between
the typical  momentum $ \sqrt{Q^2} $ transferred in the
process
and the
total
energy $\sqrt{S}$.
The most striking  phenomenological feature
at low $x$
is the rise of
the
deep inelastic cross sections.
This behaviour  is
qualitatively
expected in QCD,
and has been observed by the HERA collaborations
[\ref{DATA}], who have measured the
structure function
$F_2$
down to
$x$-values as low as $10^{-3} \div 10^{-4}$.
The observed rise
is indeed steeper than any rise one
finds
in total cross sections for soft hadronic processes. Therefore it can
only be accounted for by a hard QCD component, i.e., QCD interactions
occurring   at distance scales
much smaller than $\Lambda_{QCD}^{-1}$.
Such
scales
are controlled by perturbation theory.
  Let us
then
briefly recall
the
main
features
of QCD perturbation theory in the
region of small $x$.

Perturbative QCD predictions rely upon
the factorization theorem of collinear mass singularities [\ref{CSS}].
The use of this theorem is normally based on two approximations.
First, one picks out the
leading-twist contribution and
expresses
the
dimensionless
cross section
$F(x,Q^{2}) \sim
Q^{2} \si (x,Q^{2})$ for a
hard hadronic process as a convolution of process-dependent
coefficient functions $C_a$ and universal parton densities ${\tilde f}_a$
(Fig.~1), as follows
\begin{equation} F(x,Q^{2}) =
C_{a}(x, \as (\mu^{2}),  Q^{2}/\mu^{2}) \;
\otimes
f_{a}(x, \mu^{2})
+ {\cal O} ({ {\Lambda^2_{QCD}} / Q^2})
\;\;.
\label{F(x}
\end{equation}
Here
higher twists, i.e.
contributions from multi-parton initial states (Fig.~1b),
are neglected,
since
they are suppressed
by inverse powers  of the hard scale $Q^2$,
 and the leading-twist parton densities $f_a(x,\mu^2)$
fulfil the renormalization group evolution equations
\begin{equation}
\label{apeq}
\frac{d \;f_a(x,\mu^2)}{d\ln \mu^2} =
P_{ab}(\as(\mu^2),x)
\otimes
f_b(x,\mu^2)
\;\;,
\end{equation}
where $P_{ab} (\as ,x)$ are the generalized splitting functions,
whose
Mellin
transforms define
the
anomalous dimensions
$\gamma_{ab,N}$
\begin{equation}
\gamma_{ab,N}(\as ) \equiv \int^{1}_{0} dx \;x^{N} \;P_{ab}(\as ,x) =
P_{ab,N+1}(\as )\;\;,
\label{gamma}
\end{equation}
 $N$ being the moment conjugate to $x$.

Second,
one considers the perturbative expansion
for
the kernels
$P_{a b}$
of the $\mu^2$-evolution
\begin{equation}
P_{ab} (\as ,x) = \sum^{\infty}_{n=1} \left( \frac{\as}{2\pi}
\right)^{n} P_{ab}^{(n-1)}(x)\;\;,
\label{Pab}
\end{equation}
as well as  the  analogous expansion for the coefficients $C_a$,
and evaluates them
to
fixed order in $\as$.
Typically,
QCD perturbation theory is
at present
under control up to two loops,
that is, the terms
$P^{(0)}$ [\ref{GLAP}] and
$P^{(1)}$ [\ref{CFP},\ref{2loop}]
in the splitting functions (\ref{Pab})
are fully known, and the coefficient functions for most of the
relevant processes at
colliders
are also known to the corresponding accuracy
[\ref{Stirl}].

In the regime of low $x$, however, both the leading-twist approximation
and the fixed-order truncation of
 evolution kernels  and coefficient functions
become  critical.
On the one hand,
the leading-twist treatment violates the unitarity bound on the deep inelastic
cross section at asymptotic energies. This is related to the fact that,
as the cross section increases,
higher-twist contributions
from  parton  recombination and
rescattering
[\ref{uni1},\ref{uni2}]
become corrections of relative order $ {\cal O} (1) $
in
the
formula (\ref{F(x}). Therefore they cannot be neglected, and
 in fact  they are
essential
for asymptotic  unitarity to be restored.
On the other hand,
large
high-energy
contributions
of the type $ \as^n \ln^m 1/x$
affect
the perturbative expansions
(\ref{Pab}) (as well as  the analogous expansions for the hard
coefficients)
to all orders in $\as$.
They may spoil the convergence of the perturbative series at small $x$, as
higher powers of $\as$
associated with  multiple
hard-jet emission
may be compensated by large enhancing factors in  $\ln (1 / x)$. These
corrections have to be identified and
 summed
to all
perturbative
orders.

Throughout this paper we will not
discuss
the issue of
unitarization,
and stick to the leading-twist framework.
Phenomenological support  for this attitude is provided by the absence
of any
signal of
unitarity corrections
in
HERA data, as well as by
estimates [\ref{uni1},\ref{Dur}] showing that the onset of such corrections is
expected to be
well beyond the range of present colliders.
We will
instead
focus on the high-energy logarithmic effects in perturbation
theory.
Such effects arise
 from multiple radiation of space-like gluons in the $t$-channel
of the
process.
Each two-gluon intermediate state
(such as
that
in Fig.~2)
does
indeed contribute
a logarithmic factor
$\ln x$  for $ x  \to 0$,
or equivalently, in the moment space of Eq.~(\ref{gamma}),
a
pole $1 / N$ for $ N \to 0$.
The
resulting
higher-order structure of the
splitting functions
$P_{a b}$ is
\begin{equation}
P_{ab}^{(n-1)}(x) \sim \frac{1}{x} \left[ \ln^{n-1} x + {\cal O}(\ln
^{n-2}x) \right]  \;\;\;, \;\;\; x \to 0 \;\;.
\label{Pabn}
\end{equation}
Analogous terms show up in the coefficient functions $C_a$, but
here
we will
concentrate on the case of the
splitting functions.
The key approach to
dealing
with these potentially large
corrections is based on perturbative resummation.
 One
can consider
 improved
perturbative
expansions, which systematically sum classes of leading, next-to-leading, etc.,
small-$x$ logarithms (or $N$-poles)
to all orders in $ \as$.
For instance,
in the moment space one has the following expansion for the
anomalous dimensions
\begin{equation}
\gamma_{ab,N} {(\as)} = \sum^{\infty}_{k=1} \left[
\left(\frac{\as}{N}\right)^{k} A^{(k)}_{ab} + \as
\left(\frac{\as}{N}\right)^{k} B^{(k)}_{ab} + \dots \right] \;\;.
\label{gam}
\end{equation}
The coefficients
 $A^{(k)}$
  (leading),
 $B^{(k)}$
  (next-to-leading), etc.,
define the logarithmic hierarchy at high energy (i.e., small $x$).
Once these coefficients are known, they can be combined with expansions of the
type (\ref{Pab}) (after subtracting the resummed logarithmic terms in order to
avoid double counting), to obtain a prediction throughout the region of
$x$
where
$\as \ln (1/x)
\ltap
1 \,$
(or $\as /N
\ltap
1$),
which is much larger than the domain $\as \ln (1/x) \ll
1$ where the $\as$-expansions (\ref{Pab})
are applicable.

The outline of this paper is as follows.
In Sect.~2 we  review the present  leading-order knowledge of QCD at small
$x$, and
discuss the role of sub-leading effects. In Sect.~3 we  present the
results
of a study of next-to-leading corrections
[\ref{CHlett},\ref{CHprep}].
Sect.~4  contains the summary and some prospects.

\vskip 0.4 true cm

\noindent {\bf \boldmath 2. Power counting at small $x$}
\vskip 0.1 true cm

On the basis of the definition (\ref{gam}),
let us briefly
go through
a simple
power counting at high energy.
It is worth
recalling
first that no super-leading
terms
$\as^n/N^k ,
\;2 \, n \geq k > n$, are present in the anomalous dimensions. This effect is
associated with the coherence of gluon radiation in space-like processes at
small $x$ [\ref{BFKL}-\ref{Muel}]. Super-leading terms
do contribute to
exclusive final-state distributions, but they cancel out in inclusive
quantities.  The issue of the exclusive structure at small $x$ has  been
studied in Ref.[\ref{Ciaf}]
(some phenomenological investigations have been carried out in
Ref.[\ref{MW}]),
and will not be touched upon here.
We will then be concerned with single-logarithmic corrections at small $x$.

Note
next
that
flavour non-singlet observables do not couple to pure-gluon intermediate
states.
Therefore they are
always
less singular than $1 / x$ at small $x$,
i.e.,
in the
moment space, they are
  regular  for $N \rightarrow 0$ order-by-order in $\as$.
All the
high-energy contributions $\as^n/N^k
\;(n \geq k \geq 1)$ are thus associated
with the flavour singlet sector.

The singlet evolution equations
are
\begin{equation}
\label{sieq}
   {d \over {d \ln \mu^2}} \,
\left(\begin{array}{c}
{{\tilde f}_{S}}\\
{{\tilde f}_{g}}\\
\end{array}\right)\;
=
\left(\begin{array}{cc}
{2 N_f \ga_{q q }^{S}} & {2 N_f \ga_{q g }}\\
{ \ga_{g q }} & { \ga_{g g }}\\
\end{array}\right) \;\,
\left(\begin{array}{c}
{{\tilde f}_{S}}\\
{{\tilde f}_{g}}\\
\end{array}\right)\;
\;\;\;,
\end{equation}
where
we have used standard definitions for parton densities and anomalous
dimensions,
and
neglected
regular terms at $N \to 0$
in the quark-quark entry of the
matrix.

It is a fundamental result
which can be traced back to
the work of Lipatov and collaborators [\ref{BFKL}]
that only the gluon channel
(lower entries to the anomalous dimension matrix in Eq.~(\ref{sieq}))
contributes
to the leading order
($A^{(k)}$-type coefficients in Eq.(\ref{gam})),
and
that
the leading-order resummation
can  indeed be  carried through
by means of
an integral equation for the
 off-shell gluon Green function (BFKL equation).
The
resulting
anomalous dimensions read as follows
(${\bar \as}
\equiv C_A\as/\pi$)
\begin{equation}
\label{gad}
\ga_{gg,N}(\as)=\ga({{\abar} \over N}) + {\cal O}
(\as({\as  \over N})^k) \;\;, \;\;\;\;\;
\ga_{gq,N}(\as)=\frac{C_F}{C_A} \;
\ga_{gg,N}(\as)
+ {\cal O}(\as({\as \over N})^k) \;\;,
\end{equation}
where
the BFKL anomalous dimension $\ga(\abar/N)$ is
determined by the
implicit equation
\begin{equation}
\label{andim}
1 = \frac{\abar}{N} \; \chi (\ga({{\abar} \over N}))
\;\;, \;\; \chi (\ga)
\equiv 2 \psi(1) - \psi(\ga) - \psi(1-\ga) \;\;,
\end{equation}
$\psi$
being
the Euler $\psi$-function.
By solving Eq.~(\ref{andim})
 as a  power series in the coupling constant
one
finds
the leading $(\as/N)^k$ contributions to the gluon anomalous dimension
to all orders in $\as$. The first perturbative terms are
\begin{equation}
\label{gaper}
\ga(\abn ) = \abn + 2 \zeta(3) \left( \abn \right)^4 +
2 \zeta(5) \left( \abn \right)^6 + {\cal O}\left( \left( \abn \right)^7 \right)
\;\;,
\end{equation}
$\zeta$
being
the Riemann $\zeta$-function ($\zeta(3)\simeq 1.202,
\; \zeta(5)\simeq 1.037$).
Note that
due to
strong cancellations between real emission diagrams and virtual corrections
the coefficients of $\as^2$, $\as^3$ and $\as^5$ in Eq.~(\ref{gaper}) vanish,
and therefore
the BFKL anomalous dimension departs from its one-loop
approximation
rather slowly.
It is also worth recalling [\ref{jaro}] that
by virtue of
the
collinear
regularity of the BFKL integral kernel
the leading-order
anomalous dimensions (\ref{gad})-(\ref{andim}) are
factorization scheme invariant, i.e., they
do not depend on
the explicit procedure to regularize and factorize
 collinear mass singularities.
The physical features of the BFKL resummation may be
schematically
described considering
the asymptotic behaviour of the gluon density $f_g$. In the
fixed-coupling
limit
we find
for $ x \rightarrow 0$
\begin{equation}
\label{asyunresummed}
f_g (x, Q^2/Q_0^2) \sim \exp \left( 2 \sqrt{\abar \, \ln 1/ x \, \ln Q^2/Q_0^2}
\right) \;\;\;\; {\mbox {\rm (one-loop)}}
\end{equation}
\begin{equation}
\label{asyres}
f_g (x, Q^2/Q_0^2) \sim x^{- {\bar N}} \, \left( Q^2/Q_0^2 \right)^{1/2}
\;\;,\;\; {\bar N} \sim 4 \, \ln2 \, \abar
\;\;\;\; {\mbox {\rm (BFKL-resummed)}}
\end{equation}
 The all-order resummation of the perturbative $N$-poles (\ref{gaper})
builds up a branch point singularity
at a value $\bar N$ of the moment  proportional to $\as$.
 As a result, the gluon density increases at small $x$
like a power
rather than like $\exp ( \sqrt{ \ln (1 / x) } )$.
Also,
the associated scaling violations are stronger
in the resummed case
($(Q^2/Q^2_0)^{1/2}$ vs.
$\exp [ \sqrt{ \ln (Q^2 / Q^2_0) } ]$).
The BFKL analysis predicts a growth of the anomalous dimension
with respect to the fixed-order case,
and eventually its saturation
at
the asymptotic value $1 / 2$.

The evaluation of
 small-$x$
contributions to
the anomalous dimension matrix
 (\ref{sieq})
beyond the leading logarithmic approximation
has been the
object
of many
efforts
over the past fifteen years.
As in any perturbative expansion,
one needs to know sub-leading corrections
in order to be able to assess the
stability of the expansion and set its limits of validity.
In particular,
one may wonder whether the asymptotic
singularity
(\ref{asyres})
may change
beyond the leading order.
Moreover,
since running coupling effects mix with
sub-leading corrections,
 it is essential to know
the latter
in order to be able to carry out the renormalization group analysis
consistently
at low $x$.

It is worth
noting
that beyond the leading order
quarks start to contribute on the same footing as gluons:
corrections $ {\cal O} (\as (\as/N)^k) $
($B^{(k)}$-type coefficients in Eq.(\ref{gam}))
affect
the lower as well as the upper entries to the anomalous dimension matrix
in Eq.~(\ref{sieq}).
This
remark
is relatively trivial, but it has been
long
overlooked
in the literature,
and most studies  have focused  on the pure-gauge sector.
Let us give a perturbative  example at fixed order. Consider the
 off-shell quark Green function (Fig.~3), and its
$\as$-expansion.
In the lowest non-trivial
order ${\cal O} (\as^2)$,
the
remark
above simply
amounts
to the
well-known
fact that
the diagram in Fig.~(3b) gives rise to a {\em leading-order}
contribution  $ (\as / N) $ to the {\em gluon} anomalous dimension
(times a term $ {\cal O} (\as) $ in
the
coefficient function)
when evaluated in
the  region of ordered momenta
$ q^2 \gg k^2 \gg p^2 $,
but
it
contributes
 a {\em next-to-leading-order} correction $ \as^2 / N $ to the {\em quark}
anomalous dimension
 from the ``disordered" configuration
$ q^2 \sim k^2 \gg p^2 $.
A
mechanism
of this
kind
holds
in higher-order diagrams
as well.

Actually, from a
phenomenological viewpoint,
one may
argue that
the knowledge of the next-to-leading quark
anomalous dimensions
is even
more relevant
than that of the corresponding
corrections to the gluon anomalous dimensions. The reason for this is that
the most accurate information on small-$x \, $
QCD
is coming
from HERA data
on deep inelastic structure functions, which
couple
to quarks directly, and to gluons via a ${\cal O}(\as)$-suppressed
coefficient function.

The next-to-leading
contributions ${\cal O} ( \as ( \as / N)^k )$
to the quark  anomalous dimensions
 (upper entries to the matrix in Eq.~(\ref{sieq}))
have recently been
computed
to all orders in $ \as$
[\ref{CHlett},\ref{CHprep}]. In the next section we present the results of
this calculation.  The next-to-leading corrections to the gluon
anomalous dimensions are
in contrast
still unknown at present.  A calculational
program is however being pursued by Fadin and Lipatov [\ref{FL}].

\vskip 0.4 true cm

\noindent {\bf 3. Next-to-leading resummed results  }
\vskip 0.1 true cm

Unlike the leading-order analysis, the resummation of next-to-leading
logarithms at small $x$ is sensitive to the specific procedure one uses
to regularize and factorize the collinear mass singularities arising
from the
low-momenta  region. One then needs to develop a formalism in which
high-energy resummation
[\ref{HEF}]
is consistently matched with the all-order
collinear analysis. This can be accomplished [\ref{CHprep}]
by
virtue
of a property
of factorization at high energy, which holds for off-shell Green
functions
and supplements the one due to
the
renormalization group.

We have
performed
the
resummation
of the
next-to-leading corrections
to the quark anomalous dimensions
in two of the most commonly used factorization schemes,
  the DIS- and   the $ \msbar$-scheme.
The DIS-scheme is defined by setting the coefficient functions
for the structure function $F_2$
equal to
unity
in the quark channel and
zero
in the gluon channel, modulo an
additional condition
which is needed
to uniquely fix the gluon density to all loops
[\ref{AEM},\ref{CHprep}].
In this scheme
the resummed result
for the next-to-leading quark anomalous
dimensions
is parametrized as follows
\begin{equation}
\label{paramdis}
\ga_{qg,N}(\as)=\ga_{qg, N}^{(DIS)}(\as) + {\cal O}
(\as^2({\as  \over N})^k) \;, \;
\ga_{qq,N}^S(\as) = \frac{C_F}{C_A} \left[ \ga_{qg,N}(\as) - \frac{\as}{2\pi}
\,T_R \,\frac{2}{3} \right]
+ {\cal O}(\as^2({\as \over N})^k) \;\;,
\end{equation}
where the anomalous dimension
$\ga_{qg, N}^{(DIS)}$
can be expressed in a particularly compact form:
\begin{equation}
\label{res}
\ga_{qg, N}^{(DIS)}(\as)= \frac{\as}{2\pi} \;T_R \;\frac{2+3\ga_N-3\ga_N^2}
{3-2\ga_N} \;\frac{\Gamma^3(1-\ga_N) \,\Gamma^3(1+\ga_N)}{\Gamma(2+2\ga_N)
\,\Gamma(2-2\ga_N)} R(\ga_N)
\;\;.
\end{equation}
Here
$ \ga_N$
denotes
the BFKL anomalous dimension (\ref{andim}), and the normalization factor $R_N$
is given by
\begin{eqnarray}
\label{rn}
R(\ga_N) &=& \left\{ \frac{\Gamma(1-\ga_N) \;\chi(\ga_N)}{\Gamma(1+\ga_N)
\;[-\ga_N \,\chi^{\prime}(\ga_N)]} \right\}^{\frac{1}{2}} \nonumber \\
&\cdot& \exp \left\{ \ga_N \psi(1) + \int_0^{\ga_N} d\ga
\;\frac{\psi^{\prime}(1) - \psi^{\prime}(1-\ga)}{\chi(\ga)} \right\} \;,
\end{eqnarray}
$\chi$, $\chi^{\prime}$ being the characteristic function in
Eq.~(\ref{andim}) and its first derivative, respectively.
The resummation of the
contributions $\as(\as/N)^k$ to all orders in $\as$ is incorporated in
Eq.~(\ref{res}) through
the explicit
 $\ga_N$-dependence and
the $\as/N$-dependence of $\ga_N$ (known from the BFKL
equation (\ref{andim})).
For instance,
 using the expansion (\ref{gaper}),
one can compute
the first perturbative terms of the
anomalous dimension
(\ref{res}):
\begin{eqnarray}
\label{qadper}
\ga_{qg, N}^{(DIS)} &=& \frac{\as}{2\pi} \,T_R \,\frac{2}{3}
\left\{1 + \frac{13}{6} \abn +
\left(\frac{71}{18}-\zeta(2)\right) \left(\abn
\right)^2 +
\left[\frac{233}{27}-\frac{13}{6}\zeta(2)
+\frac{8}{3}\zeta(3)\right]
\left(\abn \right)^3
\right. \nonumber \\
&+&
\left.
\left[\frac{1276}{81}-\frac{71}{18}\zeta(2)
+\frac{91}{9}\zeta(3) - 6 \zeta (4) \right]
\left(\abn \right)^4
+
\dots
\right\}
\\
&\simeq& \frac{\as}{2\pi} \,T_R \,\frac{2}{3}
\left\{1+2.17 \abn +2.30 \left(\abn \right)^2 +8.27
\left(\abn \right)^3 +
14.92
\left(\abn \right)^4 +
\dots
\right\} \;\;.
\nonumber
\end{eqnarray}
Here
 the coefficients of the first two terms in the curly bracket agree with the
known one- and two-loop anomalous dimensions in the DIS scheme
[\ref{2loop},\ref{NDIS}], whereas
the
higher-order terms
represent new sub-leading information at small $x$.

Observe that
Eq.~(\ref{res})
does not introduce any singularity in the $N$-moment space above the leading
one
(see Eq.~(\ref{asyres})).
This means that
the position of the leading singularity is not changed by
next-to-leading
corrections
in the quark sector.
However,  the approach to it is made faster,
as a consequence of the positive sign of the corrections,
and the
resulting
scaling
violations are stronger.

It is worth noting that,
as long as
the gluon anomalous dimensions are
unknown
in next-to-leading order, the DIS-scheme may be convenient for
phenomenological investigations of the structure function $F_2$ at small $x$,
because by definition it decouples $F_2$ from gluons. The knowledge of the
quark anomalous dimensions in this scheme (Eqs.~(\ref{paramdis})-(\ref{res}))
may therefore be particularly useful.

As for  the  $\msbar $-scheme,
the resummed result
 is expressed
by relations of the type
(\ref{paramdis}) in terms of the
analogous
anomalous dimension
$\ga_{qg,N}^{({\smsbar})} (\as)$.
This is determined  implicitly
as a function
of $ \as $ and $N$
by an algebraic equation
(see Ref.[\ref{CHprep}] for full details), and
its
 perturbative expansion reads
\begin{eqnarray}
\label{qms}
\ga_{qg,N}^{({\smsbar})}
(\as) &=& \frac{\as}{2\pi} T_R \;\frac{2}{3}
\left\{ 1 + \frac{5}{3} \abn +
\frac{14}{9} \left(\abn \right)^2 +
\left[\frac{82}{81}+ 2 \, \zeta (3)    \right] \,
 \left(\abn \right)^3
+ \left[ {122 \over 243}
\right.
\right.
\nonumber\\
&+& \left.
\left.
 {25 \over 6} \, \zeta (3) \right] \,
 \left(\abn \right)^4
+
\dots
\right\}
\\
&\simeq& \frac{\as}{2\pi} T_R \;\frac{2}{3} \left\{ 1 + 1.67 \abn +
1.56 \left(\abn \right)^2 +
3.42 \left(\abn \right)^3
+
5.51 \left(\abn \right)^4 +
\dots
\right\} \;\;.
\nonumber
\end{eqnarray}
Note that the
coefficients in Eq.~(\ref{qadper}) are
systematically larger than those in Eq.~(\ref{qms}). This is
related
to the fact
that the
difference
between the two anomalous dimensions
is
proportional
to the $\msbar$-coefficients  of the structure function $F_2$,
which are quite  sizeable at small $x$ [\ref{CHprep}].

One may ask what the consequences of resummation are in terms of
the
stability
of the logarithmic  expansion at low $x$.  One must first notice that the
numerical implementation of QCD evolution
based on
resummed anomalous dimensions
is not available yet, and more work is needed. Moreover,
 a  fully consistent analysis to next-to-leading logarithmic
order obviously requires also the computation of the
gluon anomalous dimensions,
which are still unknown
to this accuracy.
  However, one can already make some
remarks
based on fixed-order features.
First, the cancellations in orders
$ {\cal O} (\as^2) $,
$\, {\cal O} (\as^3) $,
$\, {\cal O} (\as^5) $
that we observe in the leading gluon anomalous dimensions (\ref{gaper})
do no longer occur in the next-to-leading quark anomalous dimensions
(\ref{qadper}) (or (\ref{qms})).  Thus at large but finite energies
(such as
those at HERA)
we may expect the latter
to be of
comparable importance to the former
from the numerical point of view,
in spite of
their
being formally
sub-leading [\ref{CHlett}].
Second, a detailed analysis [\ref{EKL}] of the third-loop term of
the anomalous dimension (\ref{qms}) has recently
been performed,
which confirms
this expectation. The authors of Ref.[\ref{EKL}] have considered the
full two-loop evolution of parton densities
(as in standard sets of distributions,
like
the MRS' [\ref{MRS}]),
and the evolution
including the
$ {\cal O} (\as^3) $
term in eq.~(\ref{qms})
and the $ {\cal O} (\as^4) $
term in eq.~(\ref{gaper}).
They have compared the
predictions for $ F_2$ in the two cases. They have done this both in the
case of a flat input gluon distribution
at low momenta
(as
in the set MRSD$0$)
and
a steep one
(as
in the set MRSD$_{-}$). The outcome  is that  the
impact of the higher-order
corrections
heavily depends on the choice of the
input. If the input gluon
distribution
is flat, the effect may be sizeable already in the
range of $x$-values accessible at HERA. This suggests a possible instability
of the perturbative series, and calls for a careful evaluation of sub-leading
contributions to all orders.
On the other hand, a very small effect is observed in the case of a steep
input gluon: this simply means that
if the input  is more singular  than any
rise
that may ever be generated
in QCD,
then
the input
dominates the
evolution,
and the output  at large momentum scales
essentially reproduces
what
has been assumed  as an input at low scales.

\vskip 0.4 true cm

\noindent {\bf 4. Conclusions}
\vskip 0.1 true cm

QCD anomalous dimensions at small $x$
are resummed
to
leading logarithmic accuracy
by  the BFKL equation.
Sub-leading corrections
to this
approximation
are needed
for both
theoretical and phenomenological
reasons.
In this paper
we have reported on
a study
aimed at investigating
such corrections.

The
analysis
of the small-$x$ region
beyond the leading order
brings in the novel feature of
the interplay between high-energy logarithms and collinear mass singularities.
To deal with this,
a method has been set up
in which
the
small-$x$ resummation
is carried out
 in  a manner which
can unambiguously be matched
with the leading-twist
collinear factorization to all orders.

Sub-leading corrections
to the
BFKL
analysis
involve the quark and the pure-gauge sector on an equal footing.
The main achievement of the work presented here is the resummation of the
next-to-leading corrections to quark anomalous
dimensions.
Explicit results have been
obtained
in the $ \msbar $ and DIS factorization schemes.
On the other hand,
the corresponding
contributions
in the gluon sector are still unknown.

Preliminary
numerical
studies
indicate
that
the size of
the
next-to-leading effects
discussed in this paper
may be large enough to affect the deep inelastic measurements at HERA.
This
suggests
that
further and
more detailed
numerical
investigations
should be pursued
(including the full
implementation of resummed results),
as well as
further
theoretical
analyses
on
sub-leading contributions
(in this respect, the
calculation of the
next-to-leading
corrections
to gluon anomalous dimensions
represents a major challenge).
Also,
 the procedure of matching
mentioned above
between the high-energy formalism and the collinear factorization
should be used
systematically to combine
small-$x$ resummed results with finite-$x$ non-logarithmic contributions
computed in fixed-order perturbation theory.

\vskip 0.7 true cm

\noindent {\bf Acknowledgments.  }
The work presented in this paper
has been carried out in
collaboration with S. Catani.
Many useful discussions with M. Ciafaloni, G. Marchesini and B. Webber
are gratefully acknowledged.

\vskip 0.7 true cm

{\large \bf References}
\begin{enumerate}

\item \label{DATA}
      ZEUS Coll., M.\ Derrick et al., \pl{316}{412}{93};
      H1 Coll., I.\ Abt et al., \np{407}{515}{93}, \pl{321}{161}{94};
      M.\ Roco, these Proceedings; K.\ M\"{u}ller, these Proceedings.

\item \label{CSS}
      J.C.\ Collins, D.E.\ Soper and G.\ Sterman, in {\it Perturbative Quantum
      Chromodynamics}, ed. A.H. Mueller (World Scientific, Singapore, 1989)
      and references therein.

\item \label{GLAP}
      V.N.\ Gribov and L.N.\ Lipatov, Sov. J. Nucl. Phys. 15 (1972) 438,
      675; G.\ Altarelli and G.\ Parisi,
      \np{126}{298}{77}; Yu.L.\ Dokshitzer, Sov. Phys. JETP  46 (1977) 641.

\item \label{CFP}
      G.\ Curci, W.\ Furmanski and R.\ Petronzio,  \np{175}{27}{80}.

\item \label{2loop}
      W.\ Furmanski and R.\ Petronzio, \pl{97}{437}{80};
%
      E.G.\ Floratos, D.A.\ Ross and C.T.\ Sachrajda, \np{129}{66}{77}
      (E \np{139}{545}{78}), \np{152}{493}{79}; A.\ Gonzalez-Arroyo, C.\ Lopez
      and F.J.\ Yndurain, \np{153}{161}{79}; A.\ Gonzalez-Arroyo and C.\ Lopez,
      \np{166}{429}{80};
      E.G.\ Floratos, P.\ Lacaze and C.\ Kounnas,
      \pl{98}{89}{81}, 225.

\item \label{Stirl}
      See, for instance, W.J.\ Stirling, in Proceedings of the Aachen
      Conference {\it QCD -- 20 years later}, eds. P.M. Zerwas and H.A.
      Kastrup (World Scientific, Singapore, 1993), pag.~387.

\item \label{uni1}
      L.V.\ Gribov, E.M.\ Levin and M.G.\ Ryskin, \prep{100}{1}{83};
      E.M. Levin and M.G. Ryskin, Phys. Rep. 189 (1990) 268.

\item \label{uni2}
      A.H. Mueller and J. Qiu, Nucl. Phys. B268 (1986) 427;
      A.H. Mueller, Nucl. Phys. B335 (1990) 115;
%
      J.\ Bartels, preprint DESY-91-074, \pl{298}{204}{93}, Zeit. Phys.
      C60 (1993) 471;
%
      E.M.\ Levin, M.G.\ Ryskin and A.G.\ Shuvaev,
      Nucl. Phys. B387 (1992) 589;
      E.\ Laenen, E.M.\ Levin and A.G.\ Shuvaev, preprint
      Fermilab-PUB-93/243-T.

\item \label{Dur}
       A. J. Askew, J. Kwiecinski, A. D. Martin and P. J. Sutton,
       Phys. Rev. D47 (1993) 3775, preprint DTP-93-28.

\item \label{CHlett}
      S.\ Catani and F.\ Hautmann, \pl{315}{157}{93}.

\item \label{CHprep}
      S.\ Catani and F.\ Hautmann, Cambridge preprint Cavendish-HEP-94/01.

\item \label{BFKL}
      L.N.\ Lipatov, Sov. J. Nucl. Phys. 23 (1976) 338; E.A.\ Kuraev,
      L.N.\ Lipatov and V.S.\ Fadin, Sov. Phys. JETP  45 (1977) 199; Ya.\
      Balitskii and L.N.\ Lipatov, Sov. J. Nucl. Phys. 28 (1978) 822.

\item \label{Ciaf}
      M.\ Cia\-fa\-lo\-ni, \np{296}{249}{87};
%
%
      L.V.\ Gribov, Yu.L.\ Dokshitzer, S.I.\ Troyan and V.A.\ Khoze, Sov. Phys.
      JETP 67 (1988) 1303;
%
%
      S.\ Catani, F.\ Fiorani and G.\ Marchesini,
      \np{336}{18}{90}.

\item \label{Muel}
      A. H. Mueller,  \np{415}{373}{94}.

\item \label{MW}
      G.\ Marchesini and B.R. Webber, \np{349}{617}{91}, \np{386}{215}{92}.

\item \label{jaro}
       T. Jaroszewicz, Phys. Lett. B116 (1982) 291.

\item \label{FL}
      V.S.\ Fadin and L.N.\ Lipatov, \np{406}{259}{93}.

\item \label{HEF}
      S.\ Catani, M.\ Ciafaloni and F.\ Hautmann,
      \pl{242}{97}{90},  \np{366}{135}{91}, \pl{307}{147}{93}.

\item \label{AEM}
      G.\ Altarelli, R.K.\  Ellis and G.\ Martinelli,
      \np{157}{461}{79}.

\item \label{NDIS}
      E.B.\ Zijlstra and W.L.\ van Neerven, \np{383}{525}{92}.

\item \label{EKL}
      R.K. Ellis, Z. Kunszt and E. M. Levin, preprint Fermilab-PUB-93/350-T.

\item \label{MRS}
      A. D. Martin, R. G. Roberts and W. J. Stirling,
      \pl{306}{145}{93}.

\end{enumerate}







\end{document}